\documentstyle[twoside,psfig]{article}

\catcode`\@=11
\long\def\@makefntext#1{
\protect\noindent \hbox to 3.2pt {\hskip-.9pt
$^{{\eightrm\@thefnmark}}$\hfil}#1\hfill}               

\def\thefootnote{\fnsymbol{footnote}}
\def\@makefnmark{\hbox to 0pt{$^{\@thefnmark}$\hss}}    
        
\def\ps@myheadings{\let\@mkboth\@gobbletwo
\def\@oddhead{\hbox{}
\rightmark\hfil\eightrm\thepage}
\def\@oddfoot{}\def\@evenhead{\eightrm\thepage\hfil
\leftmark\hbox{}}\def\@evenfoot{}
\def\sectionmark##1{}\def\subsectionmark##1{}}

\oddsidemargin=\evensidemargin
\renewcommand{\thefootnote}{\fnsymbol{footnote}}

\newcounter{sectionc}\newcounter{subsectionc}\newcounter{subsubsectionc}
\renewcommand{\section}[1] {\vspace{12pt}\addtocounter{sectionc}{1}
\setcounter{subsectionc}{0}\setcounter{subsubsectionc}{0}\noindent
        {\tenbf\thesectionc. #1}\par\vspace{5pt}}
\renewcommand{\subsection}[1] {\vspace{12pt}\addtocounter{subsectionc}{1}
  \setcounter{subsubsectionc}{0}\noindent
  {\bf\thesectionc.\thesubsectionc. {\kern1pt \bfit #1}}\par\vspace{5pt}}
\renewcommand{\subsubsection}[1]{\vspace{12pt}%
  \addtocounter{subsubsectionc}{1}
        \noindent{\tenrm\thesectionc.\thesubsectionc.\thesubsubsectionc.
        {\kern1pt \tenit #1}}\par\vspace{5pt}}
\newcommand{\nonumsection}[1] {\vspace{12pt}\noindent{\tenbf #1}
        \par\vspace{5pt}}

\newcounter{appendixc}
\newcounter{subappendixc}[appendixc]
\newcounter{subsubappendixc}[subappendixc]
\renewcommand{\thesubappendixc}{\Alph{appendixc}.\arabic{subappendixc}}
\renewcommand{\thesubsubappendixc}
        {\Alph{appendixc}.\arabic{subappendixc}.\arabic{subsubappendixc}}

\renewcommand{\appendix}[1] {\vspace{12pt}
        \refstepcounter{appendixc}
        \setcounter{figure}{0}
        \setcounter{table}{0}
        \setcounter{lemma}{0}
        \setcounter{theorem}{0}
        \setcounter{corollary}{0}
        \setcounter{definition}{0}
        \setcounter{equation}{0}
        \renewcommand{\thefigure}{\Alph{appendixc}.\arabic{figure}}
        \renewcommand{\thetable}{\Alph{appendixc}.\arabic{table}}
        \renewcommand{\theappendixc}{\Alph{appendixc}}
        \renewcommand{\thelemma}{\Alph{appendixc}.\arabic{lemma}}
        \renewcommand{\thetheorem}{\Alph{appendixc}.\arabic{theorem}}
        \renewcommand{\thedefinition}{\Alph{appendixc}.\arabic{definition}}
        \renewcommand{\thecorollary}{\Alph{appendixc}.\arabic{corollary}}
        \renewcommand{\theequation}{\Alph{appendixc}.\arabic{equation}}
        \noindent{\tenbf Appendix \theappendixc #1}\par\vspace{5pt}}
\newcommand{\subappendix}[1] {\vspace{12pt}
        \refstepcounter{subappendixc}
        \noindent{\bf Appendix \thesubappendixc. {\kern1pt \bfit #1}}
        \par\vspace{5pt}}
\newcommand{\subsubappendix}[1] {\vspace{12pt}
        \refstepcounter{subsubappendixc}
        \noindent{\rm Appendix \thesubsubappendixc. {\kern1pt \tenit #1}}
        \par\vspace{5pt}}

\newcommand{\textlineskip}{\baselineskip=13pt}
\newcommand{\smalllineskip}{\baselineskip=10pt}

\newcommand{\publisher}[2]{{\begin{center}\footnotesize\smalllineskip
        #1
        \end{center}
        }}

\def\abstracts#1#2#3{{
        \centering{\begin{minipage}{4.5in}\footnotesize\baselineskip=10pt
        \centerline{\footnotesize ABSTRACT}
        \parindent=0pt #1\par
        \parindent=15pt #2\par
        \parindent=15pt #3
        \end{minipage}}\par}}

\def\keywords#1{{
        \centering{\begin{minipage}{4.5in}\footnotesize\baselineskip=10pt
        {\footnotesize\it Keywords}\/: #1
         \end{minipage}}\par}}

\renewenvironment{thebibliography}[1]
        {\frenchspacing
         \ninerm\baselineskip=11pt
         \begin{list}{[\arabic{enumi}]}
        {\usecounter{enumi}\setlength{\parsep}{0pt}
        \setlength{\leftmargin 13.7pt}{\rightmargin 0pt}
         \setlength{\itemsep}{0pt} \settowidth
        {\labelwidth}{[#1]}\sloppy}}{\end{list}}

\newcounter{itemlistc}
\newcounter{romanlistc}
\newcounter{alphlistc}
\newcounter{arabiclistc}

\newcommand{\fcaption}[1]{
        \refstepcounter{figure}
        \setbox\@tempboxa = \hbox{\footnotesize Fig.~\thefigure. #1}
        \ifdim \wd\@tempboxa > 5in
           {\begin{center}
        \parbox{5in}{\footnotesize\smalllineskip Fig.~\thefigure. #1}
            \end{center}}
        \else
             {\begin{center}
             {\footnotesize Fig.~\thefigure. #1}
              \end{center}}
        \fi}

\newcommand{\tcaption}[1]{
        \refstepcounter{table}
        \setbox\@tempboxa = \hbox{\footnotesize Table~\thetable. #1}
        \ifdim \wd\@tempboxa > 5in
           {\begin{center}
        \parbox{5in}{\footnotesize\smalllineskip Table~\thetable. #1}
            \end{center}}
        \else
             {\begin{center}
             {\footnotesize Table~\thetable. #1}
              \end{center}}
        \fi}

\def\@citex[#1]#2{\if@filesw\immediate\write\@auxout
        {\string\citation{#2}}\fi
\def\@citea{}\@cite{\@for\@citeb:=#2\do
        {\@citea\def\@citea{,}\@ifundefined
        {b@\@citeb}{{\bf ?}\@warning
        {Citation `\@citeb' on page \thepage \space undefined}}
        {\csname b@\@citeb\endcsname}}}{#1}}

\newif\if@cghi
\def\cite{\@cghitrue\@ifnextchar [{\@tempswatrue
        \@citex}{\@tempswafalse\@citex[]}}
\def\citelow{\@cghifalse\@ifnextchar [{\@tempswatrue
        \@citex}{\@tempswafalse\@citex[]}}
\def\@cite#1#2{{$\null^{#1}$\if@tempswa\typeout
        {IJCGA warning: optional citation argument
        ignored: `#2'} \fi}}

\def\pmb#1{\setbox0=\hbox{#1}
        \kern-.025em\copy0\kern-\wd0
        \kern.05em\copy0\kern-\wd0
        \kern-.025em\raise.0433em\box0}

\def\fnt#1#2{\footnotetext{\kern-.3em
        {$^{\mbox{\scriptsize #1}}$}{#2}}}

\font\tenrm=cmr10
\font\tenit=cmti10
\font\tenbf=cmbx10
\font\bfit=cmbxti10 at 10pt
\font\ninerm=cmr9

\font\eightrm=cmr8

\def\qed{\hbox{${\vcenter{\vbox{                   
   \hrule height 0.4pt\hbox{\vrule width 0.4pt height 6pt
   \kern5pt\vrule width 0.4pt}\hrule height 0.4pt}}}$}}

\renewcommand{\thefootnote}{\fnsymbol{footnote}}   

\begin{document}
\setlength{\textheight}{7.7truein}  


\normalsize\textlineskip
\pagestyle{empty}


\vspace*{0.88truein}

\centerline{\bf EXACT SOLUTION OF A MODEL FOR}
\baselineskip=13pt
\centerline{\bf CROWDING AND INFORMATION TRANSMISSION}
\baselineskip=13pt
\centerline{\bf IN FINANCIAL MARKETS}
\vspace*{0.37truein}

\centerline{\footnotesize R. D'HULST and G. J. RODGERS}
\baselineskip=12pt
\centerline{\footnotesize\it Rene.DHulst@brunel.ac.uk and
G.J.Rodgers@brunel.ac.uk}
\baselineskip=10pt
\centerline{\footnotesize\it Department of Mathematical Sciences, Brunel
University,}
\baselineskip=10pt
\centerline{\footnotesize\it Uxbridge, Middlesex, UB8 3PH, UK.}
\vspace*{0.225truein}
\publisher{ }

\vspace*{0.21truein}
\abstracts{An exact solution is presented to a model that mimics the crowding
effect in financial markets which arises when groups of agents share 
information. We show that the size distribution
of groups of agents has a  power law tail with an exponential
cut-off. As the size of these groups determines the supply and demand
balance, this implies heavy tails in the distribution of price variation. The
moments of the distribution are calculated, as well as the kurtosis. We
find that the kurtosis is large for all model parameter values and that the model is not self-organizing.}{}{}

\vspace*{5pt}
\keywords{Herding, economy, market organization, statistics.}

\vspace*{1pt}\textlineskip      
\section{Introduction}  
\vspace*{-0.5pt}

Empirical analyses of financial price-data show that short time variations
of the price of different assets deviate from the Gaussian
distribution\cite{mandelbrot}, which would be expected if agents were trading independently. Anomalously large oscillations are far more
probable than predicted by a Gaussian distribution and are distributed 
according to an exponentially truncated power law\cite{pictet}.

Physically motivated models used to describe this phenomena assume that it
is caused by cooperative behaviour of the traders\cite{cont,stauffer}. These models make use of the properties of system in a critical state, where the 
quantity describing the range of the interactions between agents is diverging\cite{stanley}. Generally, in a  critical state, the elements that make up the system behave synchronously and a microscopic formulation is no longer required to describe kinetics of the whole system\cite{bak}.

Having made this observation, Cont and Bouchaud\cite{cont} proposed a model of
randomly connected agents to describe crowding in financial markets. Connected agents are agents sharing the same information and making the same decision between buying, selling or doing nothing. In their original model, any pair of agents is connected with a probability $p$, corresponding to bond percolation in an infinite dimensional space\cite{stauffer2} or equivalently, to agents being vertices of a random graph\cite{bollobas}. To drive the system into
a critical state, the probability $p$ that two agents are connected has to be inversely proportional to the number of agents, $N_0$. Assuming that $p$ is proportional to $1/N_0$, they find a power law distribution for the cluster size, with an exponential cut-off, in reasonable agreement with the empirical data. When $p$ is exactly equal to $1/N_0$, the exponential correction disappears.

We present in this paper an exact solution to an extension of the Cont-Bouchaud
model\cite{cont}, introduced by Egu\'{\i}luz and Zimmermann\cite{eguiluz}. Instead of supposing that two agents are connected with a probability $p$, they considered that the network of agents is building itself and changing dynamically as decisions are made by the agents. Egu\'{\i}luz and Zimmermann\cite{eguiluz} investigated numerically the simplest case where the network of agents is built randomly. As expected, they obtained numerically the same distribution for the returns as Cont and
Bouchaud\cite{cont}.

In Sec. 2, we present the model of Egu\'{\i}luz and
Zimmermann\cite{eguiluz} and find an exact analytical expression for the
size distribution of the clusters of agents. A cluster of agents is a set of
agents making the same decision, between buying, selling and doing nothing.
We assume that the traded amount is proportional to the number of agents in
a cluster\cite{eguiluz}. Then, the distribution of returns, which is the
distribution of the relative difference between the number of buyers and
the number of sellers, is a simple function of the cluster size
distribution. In Sec. 3, a generalisation of the model is proposed.
Although the cluster size distribution is not obtained exactly in this
case, its asympotic behaviour is determined and a method to obtain its
various moments is explained.

\setcounter{footnote}{0}
\renewcommand{\thefootnote}{\arabic{footnote}}

\section{The Model}
Consider a set of $N_0$ agents in which connected agents are able to
exchange information. An agent can be connected to any of the $N_0 -1$ other agents. When it is possible to go from one agent to another following a path of connected agents, these two agents are said to belong to the same
cluster, that is, a cluster is a set of agents that can exchange information, either directly or through the intermediary of other agents. All the agents are isolated at the beginning of the simulation. At each time step, an agent is selected at random. With a probability $a$, this agent decides to buy or sell. In this case, all the agents in the same cluster follow the lead agent in buying or selling and the cluster is broken up into isolated agents. With a probability $1-a$, he decides to do nothing. In this case, two agents are selected at random and connected to each other.

When an agent decides to make a transaction, buying or selling, he triggers
the action of all the agents belonging to the same cluster. If the selected
agent is buying or selling, all the agents belonging to the same cluster
are buying or selling at the same time, respectively. Hence, the variation
of the supply and demand balance, that is, the return, is a function of the
size distribution of the clusters of agents. In this model, the return $R$
at a given time is defined as the ratio of the agents buying or selling at
this time. If the agents are buying, the return is positive and it is
negative if the agents are selling. Hence, if $n_s$ is the number of
clusters of size $s$, the probability to have a return of size $s/ N_0$ is
given by $s n_s /N_0$.

The number $n_s$ of clusters of size $s>1$ evolves like

\begin{equation}
{\partial n_s\over \partial t} = -a s n_s + {(1-a)\over N_0}
\sum_{r=1}^{s-1} r n_r (s-r) n_{s-r} - {2(1-a) s n_s\over N_0}
\sum_{r=1}^{\infty} r n_r
\label{eq:evolution of ns}
\end{equation}
where $N_0$ is the total number of agents. Note that one time step in this
formulation corresponds to one attempted update per agent in the numerical
simulation. The first term on the right hand side of Eq.~(\ref{eq:evolution
of ns}) describes the fragmentation of a cluster of size $s$ in case of a
transaction involving $s$ agents. The second term describes the creation of
a new cluster of size $s$ by coagulation of two clusters of size $r$ and
$s-r$, with $r<s$. The last term on the right hand side describes the
disappearance of a cluster of size $s$ by coagulation with another cluster.
The number of clusters of size 1, $n_1$ obeys

\begin{equation}
{\partial n_1\over \partial t} = a \sum_{r=2}^{\infty} r^2 n_r - {2(1-a)
n_1\over N_0} \sum_{r=1}^{\infty} r n_r.
\label{eq:evolution of n1}
\end{equation}
The first term on the right hand side describes the appearance of $s$
clusters of size 1 by fragmentation of a cluster of size $s$ while the second
term describes the disappearance of clusters of size 1 by coagulation with
another cluster.

When the system reaches a stationary state, the size distribution of the
clusters obeys

\begin{equation}
s n_s = {1-a\over (2-a) N_0} \sum_{r=1}^{s-1} r (s-r) n_r n_{s-r}
\end{equation}
for $s>1$ and

\begin{equation}
n_1 = {a\over 2 (1-a)} \sum_{r=2}^{\infty} r^2 n_r.
\end{equation}
Introducing the generating function

\begin{equation}
g (\omega ) = \sum_{r=2}^{\infty} r n_r e^{-\omega r},
\label{eq:generating function}
\end{equation}
we find

\begin{equation}
g (\omega ) = {(2-a) N_0\over 4 (1-a) } \left( 1 - \sqrt{1 - {4 (1-a)\over
(2-a)^2} e^{-\omega}}\right)^2,
\end{equation}
giving a size distribution for the clusters of

\begin{equation}
n_s = {(1-a)^{s-1}(2s-2)!\over (2-a)^{2s-1}s!^2} N_0.
\label{eq:solution ns}
\end{equation}
Using Stirling's formula to expand this for large $s$, we have

\begin{equation}
n_s \sim N_0 \left({4 (1-a)\over (2-a)^2}\right)^s s^{-5/2}.
\end{equation}
Hence, the model presents a power law distribution of exponent $\alpha =
5/2$ for the size of the clusters, with an exponential cut-off, as in the
model of Cont and Bouchaud\cite{cont}. For $a= 0$, the exponential
cut-off vanishes.

The properties of the cluster size distribution $n_s$ can be calculated
exactly. The number of clusters per site, $M_0$, is given by

\begin{equation}
M_0 = 2 - {(2-a)\over (1-a)}\ln (2-a).
\label{eq:number of clusters}
\end{equation}
This is easily obtained by integrating $g(\omega )$. It is interesting to note that, even if the probability of a transaction is
0, the system does not become a single cluster containing all the agents.
In fact, $a\rightarrow 0$ is the only value of $a$ that gives a power law
for the cluster size distribution. The other moments of the distribution,
$M_i$, defined as

\begin{equation}
M_i \equiv {1\over N_0} \sum_{r=1}^{\infty} r^{i} n_r,
\end{equation}
can be obtained by differentiation of the generating function. We find that
$M_1 = 1$ and $M_2 = 1/a$. As the probability of a return of size $s/ N_0$
is given by $s n_s /N_0$, the variance of the distribution of returns,
$\sigma^2$, is

\begin{equation}
\sigma^2 = M_3-M_2^2 = {(1-a)(2-a)\over a^3}.
\end{equation}
Finally, the kurtosis of the distribution of returns is given by

\begin{eqnarray}
\kappa &=& {\mu_4\over \sigma^4} - 3\\
&=& {a^4 - 18a^3 + 78a^2 - 120a + 60\over a (1-a)(2-a)}
\end{eqnarray}
where $\mu_4$ is the fourth central moment of the distribution of returns.
The kurtosis $\kappa$ is shown as a function of $a$ in Fig. 1. It
diverges at $a=0$ and $a=1$, while being asymmetric around $a=1/2$.
$\kappa$ is not small for any value of $a$, achieving a minimum value of
$\kappa \simeq 26.07$ for $a \simeq 0.846$. This shows that, even if the
cluster size distribution is not a power law for $a>0$, it has heavy tails
for every value of $a$.
\eject

\begin{figure}[htbp] 
\centerline{\psfig{file=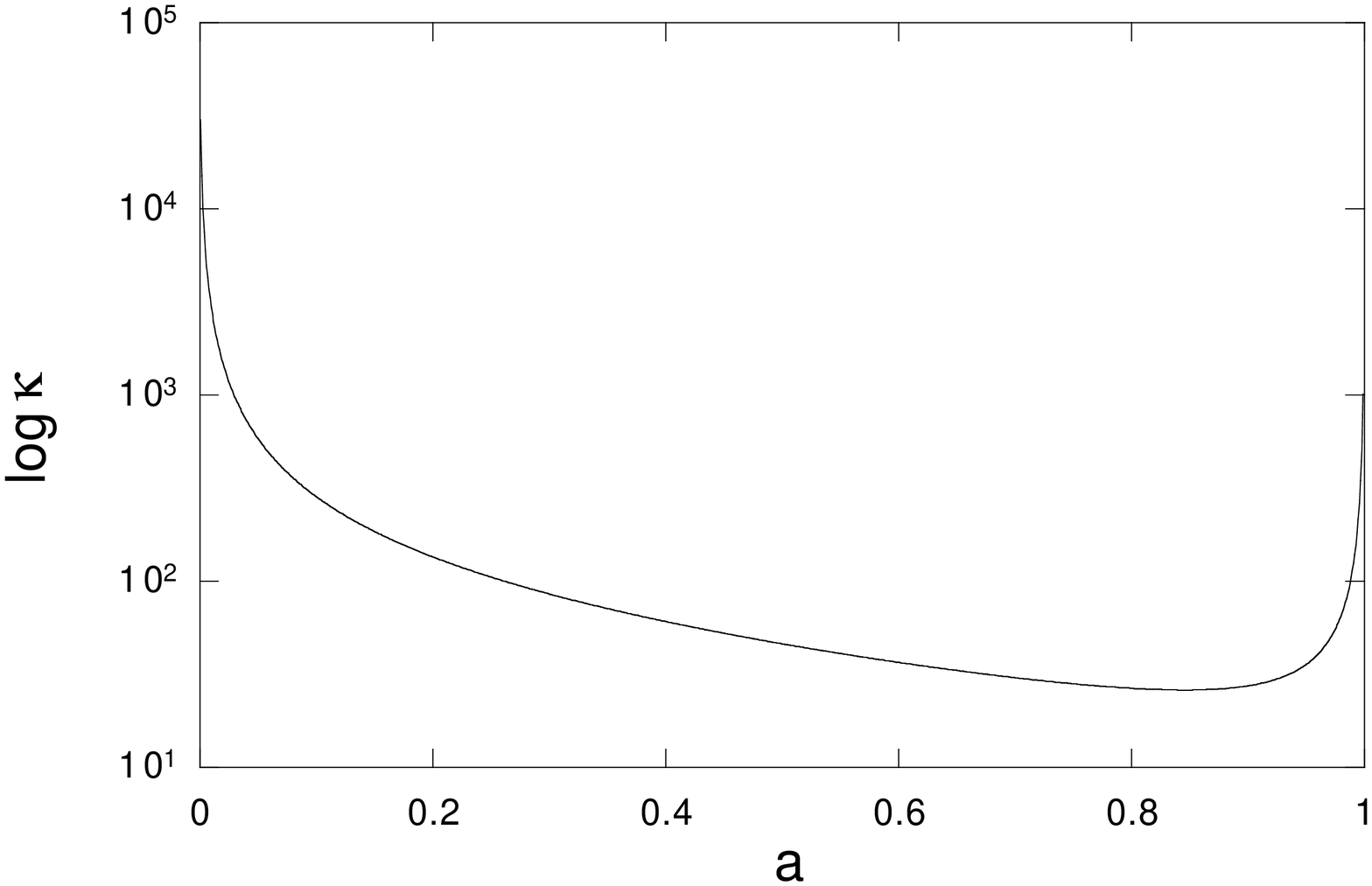,width=13cm,height=8cm}} 
\vspace*{13pt}
\fcaption{Analytical result for the kurtosis $\kappa$ of the return
distribution as a function of $a$ for $m=2$. Note that the ordinate is a
logarithmic scale.}
\end{figure}

\section{Generalisation}
The generalisation we introduce is different from the model of the previous section in that when an agent decides with a probability $1-a$ to do nothing, $m$ agents are selected at random and connected together. Then, the model of the previous section corresponds to $m=2$.

If the simulation starts with isolated agents, the possible sizes of
the clusters are given by $(m-1)i+1$, where $i$ is a non-negative integer.
It is easy to show that the combination of $m$ such clusters also gives
such a cluster. Hence, these are the only possible sizes. Excluding all the
other cluster sizes, the time evolution of the number of clusters of size
$s>1$ is given by

\begin{eqnarray}
\nonumber
{\partial n_s\over \partial t} &=& -a s n_s + {(1-a)\over N_0^{m-1}}
\sum_{i=1}^{m}\sum_{r_i=1}^{s-m+1} \left(\prod_{i=1}^{m} r_i n_{r_i}\right)
\delta_{s,\sum_{i=1}^{m}r_i}\\
&-& {m(1-a) s n_s\over N_0^{m-1}} \left(\sum_{r=1}^{\infty} r n_r\right)^{m-1}
\label{eq:generalisation evolution of ns}
\end{eqnarray}
with a particular equation for $s=1$,

\begin{equation}
{\partial n_1\over \partial t} = a \sum_{r=2}^{\infty} r^2 n_r - {m(1-a)
n_1\over N_0^{m-1}} \left(\sum_{r=1}^{\infty} r n_r\right)^{m-1}.
\label{eq:generalisation evolution of n1}
\end{equation}
When the system has reached a stationary state, the cluster size
distribution is the solution of

\begin{equation}
s n_s = {(1-a)\over (a+m(1-a)) N_0^{m-1}}
\sum_{i=1}^{m}\sum_{r_i=1}^{s-m+1} \left(\prod_{i=1}^{m} r_i n_{r_i}\right)
\delta_{s,\sum_{i=1}^{m}r_i}
\label{eq:stationary general}
\end{equation}
with

\begin{equation}
n_1 = {a\over m (1-a)} \sum_{r=2}^{\infty} r^2 n_r
\end{equation}
where the Kronecker delta $\delta_{\alpha , \beta} = 1$ when $\alpha =
\beta$ and 0 when $\alpha \not = \beta$.

Considering $m=2$, the equation giving the stationary distribution can also
be written

\begin{equation}
s n_s =\left({1-a\over (2-a) N_0}\right)^2 \sum_{r_1=1}^{s-2}
\sum_{r_2=1}^{s-1-r_1} r_1 r_2 (s-r_1-r_2) n_{r_1} n_{r_2} n_{s-r_1-r_2}.
\end{equation}
by inserting the expression for $r n_r$ into the expression of $s n_s$.
Consequently, when $s$ is large, the stationary distributions for $m=2$ and
$m=3$ display the same dependence on $s$. Introducing iteratively the
expression for $r n_r$ in $s n_s$, it is easy to show that the stationary
distribution displays the same dependence on $s$ for every value of $m$
when $s$ is large. From the previous section, the stationary distribution
is a power law with an exponential cut-off. The exponent is $\alpha = 5/2$,
independent of $m$. Considering $s n_s$ for the first values of $s$, as
obtained by Eq.~(\ref{eq:stationary general}), the cluster size
distribution also displays a dependence like

\begin{equation}
s n_s \sim N_0 \left({(1-a)\over (a+m(1-a)) }\right)^{(s-1)/(m-1)}
\left({n_1\over N_0}\right)^{s}.
\end{equation}
By iteration, $s n_s$ follows this relation for all values of $s$, giving
an exponential correction to the power law, as expected. Fig. 2 compares
numerical results for $s n_s$ for $m=2$ and $m=3$, using
Eq.~(\ref{eq:stationary general}) iteratively, with $a=0$.  An exponent of
$5/2$ is obtained in both cases. Note that in direct simulations of the
model, it is difficult to appreciate the departure from the power law from
the numerical simulations of the model for $a<0.5$, as the largest values
of $s$ are restricted. A power law fit to the results from direct numerical
simulations of the model for $m=2$, $N_0 = 10^4$ and $a=0.01$ gives an
exponent\cite{eguiluz} $\alpha \sim 2.7$. Hence, it is necessary to
consider much larger systems to make a correct estimate of $\alpha$.

\begin{figure}[htbp] 
\centerline{\psfig{file=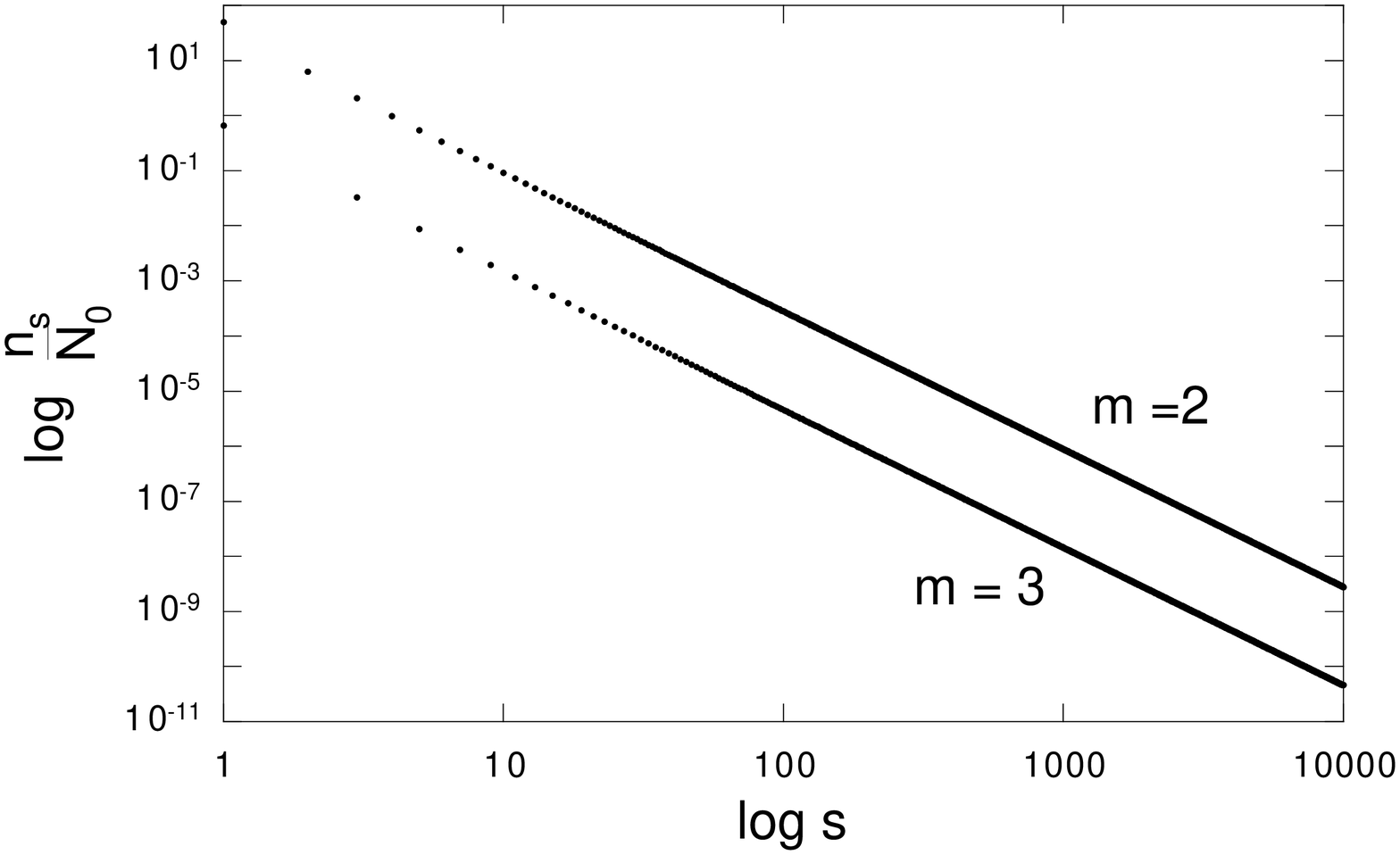,width=13cm,height=7.5cm}} 
\vspace*{13pt}
\fcaption{Numerical results for the cluster size distribution $n_s$ for
$m=2$ (upper dots) and $m=3$ (lower dots) obtained by iterating
Eq.~(\ref{eq:stationary general}) with $a = 0$. For $m=2$, $n_s$ has been
rescaled by a factor of 100 to separate the two sets of results.}
\end{figure}
Introducing the same generating function $g (\omega )$,
Eq.~(\ref{eq:generating function}), we obtain the self-consistent equation

\begin{equation}
g (\omega ) = {(1-a)\over (a+m(1-a)) N_0^{m-1}} \left( g (\omega ) + n_1
e^{-\omega } \right)^m
\label{eq:generalised generating function}
\end{equation}
with

\begin{equation}
n_1 = N_0 { a + (m-1)(1-a)\over a + m(1-a)}.
\end{equation}
By differentiating Eq.~(\ref{eq:generalised generating function}), all the
moments of $g(\omega )$ can be calculated exactly. For instance,

\begin{eqnarray}
M_2 &\equiv& {1\over N_0} \sum_{r=1}^{\infty} r^2 n_r\\
&=& {a + (m-1)(1-a)\over a}.
\end{eqnarray}

\section{Conclusions}
We have presented the exact solution to a model that mimics the herding effect
in financial market as well as the information transmission. A power law
distribution with an exponential cut-off is found for the size distribution
of the clusters of agents and hence for the distribution of returns. 
The model is not self-organized as the parameter
$a$ describing the probability of a transaction has to be tuned to a low
value to ensure the appearance of transactions of all sizes. Except for the
small values of $a$, the system does not display a power law distribution
for the large transactions. Nevertheless, we computed the kurtosis to show
that the model has heavy tails for any value of $a$. Hence, the model
deviates from Gaussian behaviour and the assumption of independent agents.

A simple generalisation of the model was also investigated analytically,
considering the number of agents connected at a time step as a free
parameter $m$. The large $s$ dependence of the distribution of the clusters
of agents is determined. The cluster size distribution is found to be
independent of $m$ in the large $s$ limit. A method to calculate exactly
any moment of the distribution was presented.

To incorporate more realistic features, several extensions of this model
are possible, such as a variable $m$ or a different process of cluster fragmentation, allowing some agents to remain connected after a transaction. However, the modifications to the equations seems to be irrelevant for large values of $s$. Consequently, the same asymptotic behaviour with a power law of exponent $\alpha = 5/2$ for the cluster size distribution is expected.

\nonumsection{References}


\begin{thebibliography}{000}
\bibitem{mandelbrot}
B. Mandelbrot, Journal of Business {\bf 36} (1963) 394--419.

\bibitem{pictet}
O. V. Pictet, M. Dacorogna, U. A. Muller, R. B. Olsen and J. R. Ward,
Journal of Banking and Finance {\bf 14} (1997) 1189--1208.

\bibitem{cont}
R. Cont and J.-P. Bouchaud, {\it Macroeconomic Dynamics} in press, preprint cond-mat/9712318.

\bibitem{stauffer}
D. Stauffer, P. M. C. de Oliveira and A. T. Bernardes, Int. J. Theor. Appl. Finance {\bf 2} (1999) 83--94; D. Chowdhury and D. Stauffer, Eur. Phys. J. B. {\bf 8} (1999) 477--482 (cond-mat/9810162). 

\bibitem{stanley}
H. E. Stanley, {\it Introduction to Phase Transitions and Critical Phenomena}, Oxford University Press, New York (1971).

\bibitem{bak}
P. Bak, {\it How nature works: the science of self-organized criticality}, Oxford University Press, Oxford (1997).

\bibitem{stauffer2}
D. Stauffer and A. Aharony, {\it Introduction to Percolation Theory}, Taylor and Francis, London (1994).

\bibitem{bollobas}
B. Bollobas, {\it Modern Graph Theory}, Springer, New York (1998).

\bibitem{eguiluz}
V. M. Egu\'{\i}luz and M. G. Zimmermann, preprint cond-mat/9908069.
\end{thebibliography}
\end{document}